%% file: ms.tex
\documentclass[12pt,preprint]{aastex}
\usepackage{natbib}
\usepackage{epsfig}

\def\cmtwo{\hbox{cm$^{-2}$}}
\def\ltsim{$\lesssim$}
\def\cc{\hbox{cm$^{-3}$}}
\def\kms{\hbox{km s$^{-1}$}}
\def\deeg{\hbox{$^{\rm o}$}}
\def\HI{\hbox{${\rm H^o}$}}

\def\MgII{\hbox{${\rm Mg^+}$}}
\def\NaI{\hbox{${\rm Na^o}$}}
\def\CaII{\hbox{${\rm Ca^+}$}}
\def\FeII{\hbox{${\rm Fe^+}$}}
\def\HeI{\hbox{${\rm He^o}$}}

\def\glong{\hbox{$l^{\rm II}$}}
\def\glat{\hbox{$b^{\rm II}$}}
\def\dVi{\hbox{$dV_{\rm i}$}}
\def\dV{\hbox{$dV$}}
\def\NHI{\hbox{$N{\rm (H^o)}$}}
\def\NMgII{\hbox{$N{\rm (Mg^+)}$}}
\def\NCaII{\hbox{$N{\rm (Ca^+)}$}}
\def\Lymanalpha{\hbox{L$\alpha$}}
\def\VAC{\hbox{$V$(AC)}}
\def\VBF{\hbox{$V$(BF)}}

\def\ACV{\hbox{--35.1$\pm$0.6}}
\def\ACl{\hbox{12.7}}
\def\ACb{\hbox{14.6}}
\def\BFV{\hbox{--28.1$\pm$4.6}}

\def\HeV{\hbox{--26.3$\pm$0.4}}
\def\Hel{\hbox{3.6}}
\def\Heb{\hbox{15.3}}

\slugcomment{Submitted to the Astrophysical Journal}

\shorttitle{LISM Velocity Distribution}
\shortauthors{Frisch}

\received{2003 February 1}
\begin{document}
\title{Local Interstellar Matter:  The Apex Cloud }

\author{Priscilla C. Frisch}
\affil{University of Chicago, Department of Astronomy and Astrophysics, 5460 S. Ellis Avenue, Chicago, IL 60637}

\begin{abstract}

Several nearby individual low column density interstellar cloudlets
have been identified previously based on kinematical features evident in
high-resolution \CaII\ observations near the Sun.  One of these
cloudlets, the `Apex Cloud'' (AC), is within 5 pc of the Sun in the
solar apex direction.
The question of which interstellar cloud will constitute the next
galactic environment of the Sun in principle can be determined from
cloudlet velocities.  The interstellar absorption lines towards $\alpha$ Cen
(the nearest star) are consistent within measurement uncertainties with the
projected G-cloud (GC) and AC velocities, and also with the velocity of the cloud inside of 
the solar system (the Local Interstellar Cloud, LIC)
providing a small velocity gradient is present in the LIC.
The high GC column density towards
$\alpha$ Oph compared to $\alpha$ Aql suggests that  $\alpha$ Aql
may be embedded in the GC so that the AC would be closer to the Sun
than the GC.  This scenario favors the AC
as the next cloud to be encountered by the Sun, and the AC would have a supersonic
velocity with respect to the LIC.
The weak feature at the AC velocity towards 36 Oph suggests that the AC
cloud is either patchy or does not extend to this direction.
Alternatively, if the GC is the cloud which is foreground to $\alpha$ Cen,
the similar values for \NHI\ in the GC components towards $\alpha$ Cen
and 36 Oph indicates this cloud is entirely contained within the nearest
$\sim$1.3 pc, and the \CaII\ GC data towards $\alpha$ Oph would then
imply a cloud volume density of $\sim$5 \cc, with dramatic consequences
for the heliosphere in the near future.


\end{abstract}


\section{Introduction}

The interstellar material (ISM) within $\sim$30 parsecs of the Sun is
one of the few regions of the Milky Way Galaxy where there is a
reasonable possibility of separating the effects of topology and
kinematics, and also identifying the properties of sub-parsec sized
cloudlets. Velocities of interstellar absorption lines seen towards
nearby stars tag cloudlets, and give advance warning of variations in
the solar galactic environment.  Interstellar material inside of the
solar system was first identified by a weak \Lymanalpha\
interplanetary glow from the resonance fluorescence of solar
\Lymanalpha\ emission scattering from interstellar \HI\ inside of the
solar system \citep[e.g.][]{BertBlam:1971,ThomasKrassa:1971}.  The
astronomical identification of this cloud was difficult, however,
since $Copernicus$ satellite observations of the interplanetary
Ly$\alpha$ glow showed that nearby ISM towards Scorpius and Ophiuchus
stars has a different velocity than the interstellar \HI\ inside of
the solar system \citep{AdamsFrisch:1977}.  Presently, the best
determination of the velocity vector of ISM inside of the solar system
(or the Local Interstellar Cloud, LIC) is derived from direct Ulysses
observations of interstellar \HeI\ inside of the solar system, and
EUVE observations of the \HeI\ 584 \AA\ backscattered emission
\citep[][Table 1]{Witteetal:2003,Flynne:1998}.  These data give a LIC
velocity of --26.3$\pm$0.4 \kms\ (all velocities here are heliocentric
unless specifically identified as in the local standard of rest, LSR).
The heliocentric upstream direction is \glong=3.4$\pm$0.5\deeg,
\glat=15.9$\pm$0.5\deeg.  This corresponds to an LSR velocity vector
of --15.8 \kms\ from an upstream direction \glong=346\deeg,
\glat=0.2\deeg.
\footnote{For converting the derived heliocentric vectors to the Local
Standard of Rest (LSR), the solar apex motion based on $Hipparcos$
data is used, corresponding to a velocity $V$=13.4 \kms, towards
\glong=27.7\deeg\ and \glat=32.4\deeg\ \citep{DehnenBinney:1998}.}
Other cloudlets besides the LIC are close to the Sun.  Interstellar
absorption line velocities towards $\alpha$ CMa (2.7 pc) and $\alpha$
Aql (5 pc) give evidence for about one cloud per 1.5 pc
\citep[e.g.][]{Lallementetal:1995}.  The Sun moves with respect to the
LIC at $\sim$5 pc per million years, and should enter a new cloud
within the next $\sim$200,000 years.  This letter explores which of
these cloudlets within 5 pc of the Sun is likely to form the next
galactic environment of the Sun.

The first step in understanding the small-scale structure of nearby
ISM requires using cloud kinematics to identify individual cloudlets.
Optical \CaII\ lines show nearby ISM flows past the Sun with an
upstream direction indicating an origin associated with the
Scorpius-Centaurus Association, and discrete cloudlets are identified
in this flow
\citep[e.g.][LVF86]{Frisch:1981,Crutcher:1982,Bzowski:1988,FrischYork:1986,Crawford:1991,Frisch:1995,Vallergaetal:1993,LallementVidalMadjarFerlet:1986}.
The \CaII\ studies are heavily weighted towards gas in the upstream
direction (the galactic hemisphere) since interstellar \CaII\ is
extremely difficult to observe towards the downstream stars because of
low column densities.  The velocity difference between upstream
interstellar gas and the LIC developed into a paradigm for describing
the velocities of nearby ISM in terms of two clouds, the 'G' Cloud
(GC) towards the galactic center hemisphere, and the LIC
\citep[][LB92]{Lallementetal:1990,LallementBertin:1992}.  The GC (with
heliocentric velocity --29.4$\pm$0.5 \kms) was derived from
observations of a group of stars between \glong=200\deeg $\rightarrow$
60\deeg, \glat=--70\deeg $\rightarrow$ 40\deeg\
\citep{Lallementetal:1990}, however many stars in this region do not
show components at the GC velocity \citep{Crawford:2001}.

An analysis of $\sim$100 interstellar \CaII\ and ultraviolet
absorption components observed towards 60 nearby stars indicates that
nearby ISM consists of a group of cloudlets flowing with bulk flow
velocity \VBF=\BFV\ \kms, from an upstream direction towards the
Scorpius-Ophiuchus Association \citep[][Paper I, Table
\ref{tab:vec}]{Frischetal:2002,Frisch:1995}.  In Paper I, we
numerically filtered data in the data set used for the analysis in
order to minimize the possibility of misidentified circumstellar
components.  GC components were subsumed in the bulk flow vector,
rather then identified as a single distinct cloud.  Many of the
absorption components were found to be grouped into individual (5--7)
kinematical and spatial groups in the rest frame of the bulk flow
(reminiscent of the LVF02 cloudlets).  One of these groups, identified
as the ``Aql-Oph'' cloud in Paper I and as ``Panoromix'' in LVF86, is
close to the solar apex motion and is here termed the ``Apex Cloud''
(AC).  The AC, seen towards $\alpha$ Aql, is close to the Sun and a
candidate for the next interstellar cloud to be encountered by the
Sun.



The analysis method used here is discussed in Paper I.  The data 
are from Paper I, supplemented by \MgII\ velocity data from
\citet[][RL02]{RedfieldLinsky:2002}, including the addition of data on
the stars $\xi$ BooA, $\chi 1$ Ori, $\kappa^1$ Cet, AU Mic, HZ 43, 101
Tau, and $\gamma$ Dra.

\section{Clouds}
\subsection{Apex Cloud \label{ac}}

In Paper I, the AC cloud was identified towards 5 stars, $\alpha$ Aql,
$\alpha$ Oph, $\zeta$ Aql, $\gamma$ Oph, and $\lambda$ Aql, and
possibly towards $\delta$ Cyg \citep[although the component is cold,
$b$(\NaI)=0.42 \kms, unlike other nearby components][]{WeltyNa:1994}.
Since the AC must be within 5 pc, it is reasonable to ask if it is
also in front of $\alpha$ Cen.  This possibility is tested here by
rederiving the AC velocity vector, using the original 5 stars sampling
the Aql-Oph cloud and the \MgII\ line velocity in $\alpha$ CenA,B
(RL02).  The resulting best-fit vector, \VAC, shows that the $\alpha$
CenA,B and the Aql-Oph components have a velocity consistent with a
single velocity vector.  A component was deemed consistent if
$|$\dVi$|$\ltsim 1.3 \kms, where \dVi\ is the difference between the
observed component velocity (not including uncertainties) and the
projected best-fit \VAC\ velocity for star i.  In an iterative
process, this best-fit velocity vector was then used to search for
other consistent velocity components with \dVi\ltsim 1.3 \kms, and
then rederived.  The final best-fit vector (Table 1) is based on 
15 stars in the galactic center hemisphere, including the --35 \kms\
component towards 36 Oph (see below).  The fit is robust in the sense that
the omission of either 36 Oph or two second quadrant (90\deeg--180\deeg) 
AC components from the fitted sample 
yields a best fit AC vector differing by $<$0.2 \kms and $<$0.3\deeg\ from the 
quoted value. The LSR AC vector corresponds to a velocity
--23.3$\pm$0.5 \kms, from the LSR upstream direction \glong=5.5\deeg,
\glat=4.1\deeg.  The heliocentric AC vector derived here is close to
the value found for the ``Panoromix'' cloud in LVF86.  Table 2 lists
the components with $|$\dVi(AC)$|$\ltsim 1.3 \kms\ for stars in the
galactic center hemisphere.  Fig. \ref{fig:1a} shows that stars with
$|$\dVi(AC)$| <$1.3 \kms\ at the AC velocity (circles) are found
predominantly in or close to the galactic center hemisphere, although
AC components in stars close to the shaded band (where
$|$\dVi(AC)--\dVi(G)$| <$2 \kms) may also be attributed to the GC
(next section).  The AC component (--19.6 \kms) towards $\alpha$ Pav (56 pc)
is relatively cold,  $T <$1,500 K with $b{\rm (\CaII)}  =$0.8 \kms\
\citep{CrawfordLallementWelsh:1998}.  Cold components have not been 
observed in closer stars, and a serendipitous coincidence with the AC velocity
can not be ruled out for any particular star, particularly more distant stars.
The --35 \kms\ component towards 36 Oph is observed only as a weak broad \HI\
feature, which \citet{WoodLinskyZank:2000} attributed to an astrosphere around 
36 Oph.  The AC contribution, if present, would be blended with the astrosphere
contribution, if real.

\subsection{``G'' Cloud \label{gc}}

Although \CaII\ data show that components at the GC velocity are
intermittent towards stars with adjacent sightlines
\citep{Crawford:2001}, and the bulk flow of ISM past the Sun is
consistent with a larger object (or shell) fragmented into cloudlets
or filaments (Paper I), the assumption is made here that the GC is
real.  (The alternative assumption is that the GC components are a
subset of the bulk flow vector, with a serendipitous coincidence in
velocity.)  The GC vector is then rederived in the same manner as the
AC vector by fitting a best-fit vector through components in the stars
$\alpha$ CenA,B, 36 Oph, $\alpha$ Aql, 70 Oph, and $\alpha$ Oph, and
then searching for matching components in galactic center hemisphere
stars.  These matching components were refit for a better vector,
yielding an LSR GC velocity of --17.7$\pm$0.4 \kms\ from an upstream
direction \glong=351.1\deeg, \glat=+8.5\deeg.  The heliocentric GC
vector derived here is very close to the LB92 G vector
(Table 1).  The bulk velocity vector ($V$(96), Table 1) of nearby
cloudlets flowing past the Sun is also close to the G-cloud vector, since
both derivations are based primarily on stars in the upstream
direction where higher column densities make ground data more
available.  Stars showing components at the projected GC velocity (to
within $|$\dVi(GC)$|<$1.3 \kms) are plotted as crosses in
Fig. \ref{fig:1a}.

\subsection{LIC}

For comparison, stars showing showing components at the projected LIC
velocity, with $|$\dVi(LIC)$|<$1.3 \kms, are plotted as crosses in
Fig. \ref{fig:1b}.  
The LIC is generally well-defined in the downstream hemisphere (\glong=90\deeg--270\deeg).
There is cluster of components at the LIC
velocity, but probably unrelated to the LIC, near \glong$\sim$40\deeg, 
\glat=20--70\deeg.  The stars showing these components are in the 
tangential region of the Upper Centaurus-Lupus shell of the Scorpius-Centaurus superbubble, which has expanded to the solar vicinity
\citep{Frisch:1981,Crawford:1991,deGeus:1992,Frisch:1995}.  The nearest star in
the group, HD 131156 (\glong=23\deeg, \glat=61\deeg), is 7 pc away.

\section{Discussion \label{disc}}

\subsection{$\alpha$ Cen}

One test of whether the AC or GC best describes the motion of the
nearest upstream ISM constituting the future galactic
environment of the Sun depends on the ISM component towards the
nearest star $\alpha$ Cen.  The GC fits the $\alpha$ CenAB component
velocities best (\dV=--0.1, --0.4 \kms), although the AC also
provides an excellent fit (\dV=0.6, 0.3 \kms).
The apparent absence of a component at the
LIC velocity towards $\alpha$ Cen has been a longstanding puzzle
\citep{Landsmanetal:1984}.  Hubble Space Telescope observations of
\MgII\ and \FeII\ give a weighted mean interstellar gas velocity of
--18.0$\pm$0.5 \kms\ towards
$\alpha$ CenAB, and a range --16.2 to --18.5
\kms\ when maximum reported measurement uncertainties are included (RL02).
The projected LIC velocity vector towards $\alpha$ CenAB is --16.7 to
--17.2 \kms, and the difference between the projected LIC velocity
and the observed velocity (\dV$<$1.0 \kms) is less than the reported
\citep{LinskyWood:1996} 
Doppler $b$-value (2.3 \kms), turbulent line broadening 
($\sim$1.2 \kms),
and nominal 3 \kms\ instrumental resolution.  The data are consistent
with the LIC 
filling the $\alpha$ Cen sightline, provided a small velocity
gradient or discontinuity (e.g. $\simeq$0.5--1 \kms) is present in the cloud.
The original conclusion that the upstream ISM velocity differs
from the ISM velocity observed inside of the solar
system was based on observations of the weak Ly$\alpha$ 1215 A glow
from solar radiation fluorescing on interstellar \HI\ inside of the
solar system \citep{AdamsFrisch:1977,Lallementetal:1990}.  However,
the radiation pressure required to interpret the interplanetary
Ly$\alpha$ glow was unknown for these early data, introducing large
uncertainties in the analysis of the Ly$\alpha$ glow data.  The
improved \HeI\ LIC velocity reduced the discrepancy with the
$\alpha$ Cen data.
The small difference between the cloud temperature
towards $\alpha$ Cen \citep[5,400$\pm$500,][]{LinskyWood:1996} and the
LIC \citep[6,300$\pm$340 K,][]{Witteetal:2003} is 
consistent with small temperature gradients predicted by
radiative transfer models for the LIC \citep{SlavinFrisch:2002}.
The GC and AC clouds both provide an acceptable fit to the
$\alpha$ Cen, and the GC is the closest.  The LIC also provides an acceptable
fit if a small velocity gradient or discontinuity is present in the cloud.
The explanation that the observed broadening of the $\alpha$ Cen Ly$\alpha$ absorption line
due to the heating of \HI\ in the outer heliosheath remains viable
\citep{LinskyWood:1996,Gayleyetal:1997}.

\subsection{The Apex Cloud versus G-Cloud as Future Environment of Sun}

\begin{figure}
\vspace*{4.in}
\includegraphics{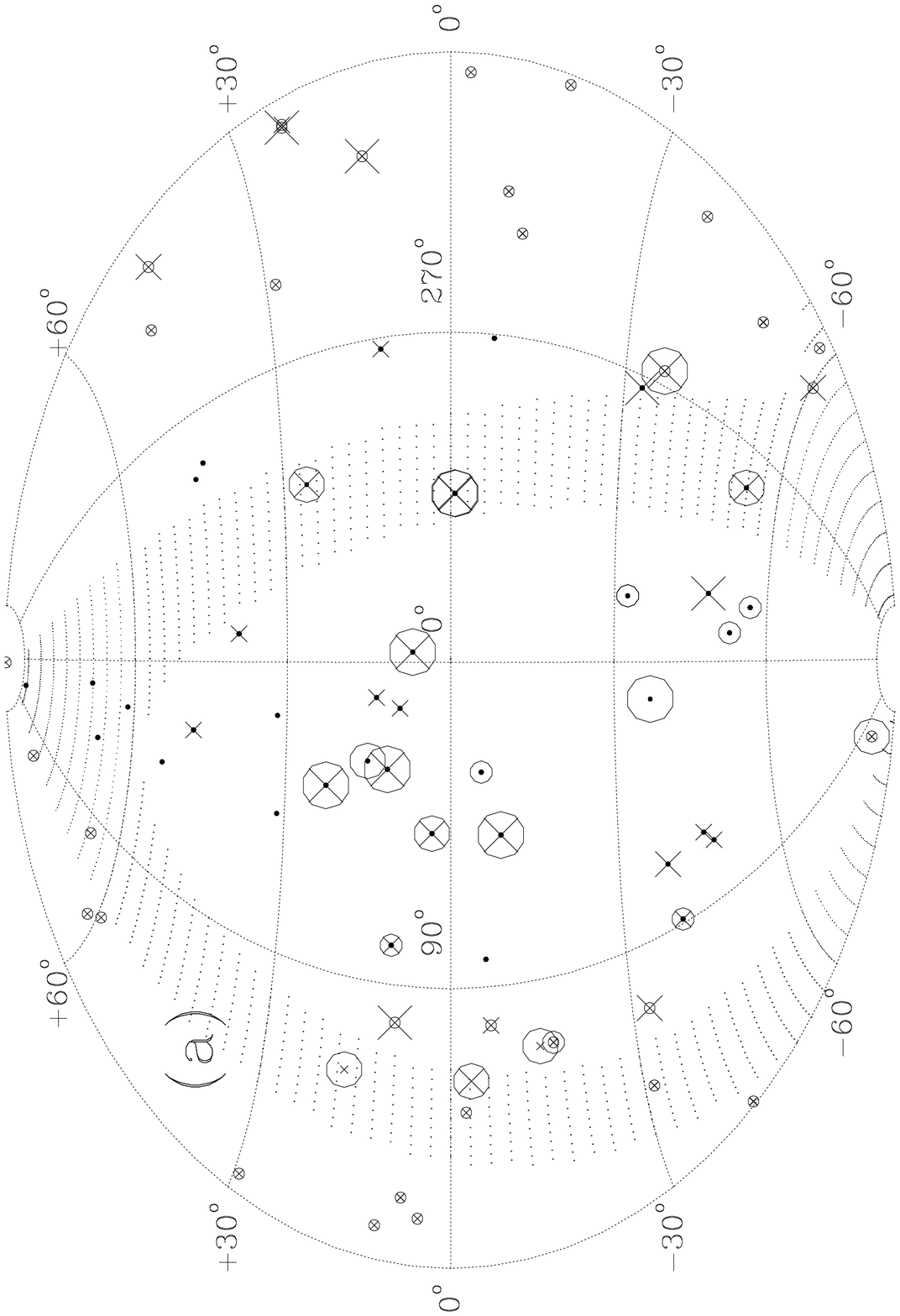}
\caption{(a) Locations of stars showing a velocity absorption
component with $|$\dVi(AC)$| <$1.3 \kms (circles) or
$|$\dVi(GC)$|<$1.3 \kms (crosses).  The regions where
$|$\dVi(AC)--\dVi(GC)$| <$2 \kms\ are shaded.  The size of the symbol
is inversely related to the star distance.  The plain dots show stars
with no components at the AC or GC velocities.
\label{fig:1a} }
\end{figure}

Since the current observational uncertainties do not allow the identity of the cloud towards $\alpha$ Cen AB to be firmly determined,
$\alpha$ Cen can not be used to determine the next interstellar cloud
to be encountered by the solar system.  The stars $\alpha$ Aql, 36 Oph, 
70 Oph, and $\alpha$ Oph provide the next set of tests as to whether
the AC or GC (if it is real) will form the future solar environment.
Fig. \ref{fig:1a} shows that the nearest stars in the galactic center hemisphere tend to show
components at both the GC and AC velocities, while more distant stars
tend to show only the GC.  Both the AC and GC are candidates for the
next encountered cloud.

Component properties may help select whether the AC or GC is likely to
be closer to the Sun.  The high GC column density towards $\alpha$ Oph
compared to $\alpha$ Aql suggests that $\alpha$ Aql is embedded in the GC,
and that the AC is closer to the Sun than the GC.  
Towards 70 Oph the
GC/AC ratio of \NMgII\ is $\sim$23 (RL02), while the GC/AC ratios for
\NCaII\ towards $\alpha$ Aql and $\alpha$ Oph are, respectively, 3 and
10.  Thus the GC has significantly larger column densities than the AC.
Both \MgII\ and \CaII\ are enhanced in the gas phase by grain destruction, so the
larger GC column densities may signify a large cloud length, grain destruction \citep{Frischetal:1999},
or high volume densities.  However, since $\alpha$ Cen and 36 Oph have comparable
column densities \citep[to within $\sim$15\%,][]{LinskyWood:1996,WoodLinskyZank:2000},
the large cloud length possibility is not viable.
The extents of the AC and GC are shown in Fig. \ref{fig:1a}, indicating both clouds
extend to the solar apex direction.  Both clouds are patchy, and both
clouds have an LSR upstream direction $\sim$40\deeg\ away from the
solar apex direction (Table 1).  Towards 36 Oph the AC component is
barely detected, and was attributed originally to an astrosphere
around 36 Oph \citep{WoodLinskyZank:2000}.  Towards $\alpha$ Oph, the
GC is extremely patchy \citep{Crawford:2001}, which was originally
attributed to the patchy \HI\ 21 emission found at this velocity and
in this region \citep{FrischYorkFowler:1987}.

\begin{figure}
\vspace*{4.in}
\includegraphics{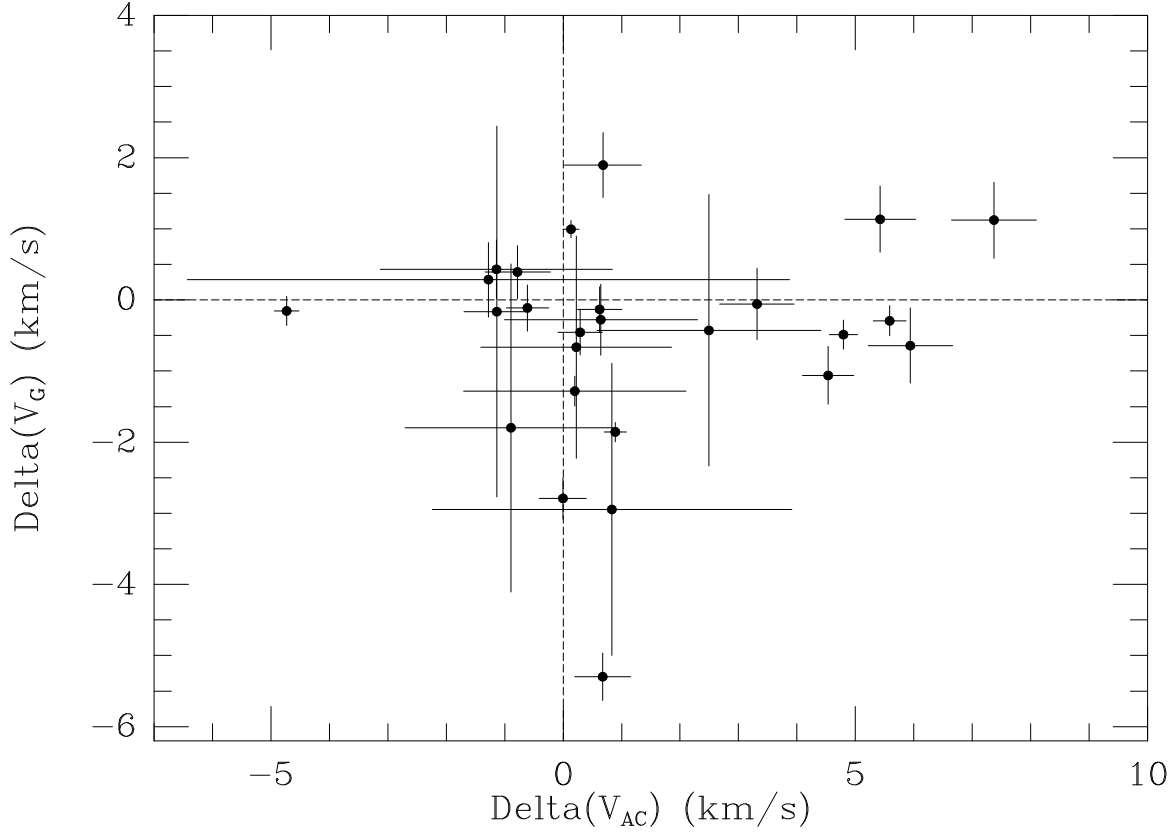}
\caption{(a) This figure displays the ability of the AC and GC vectors
to describe components seen towards each star in the galactic center
hemisphere.  The minimum \dVi(AC) (x-axis) and the minimum \dVi(AC)
(y-axis) are plotted for each star.  Where only one absorption
component is present, \dVi(AC) and \dVi(GC) are calculated for the
same component.  Where several components are present for star i, the
component with the minimum \dVi(AC) value may be different from the
component showing the minimum \dVi(GC) value.  If either the AC (or
GC) vector perfectly described the velocity components in the galactic
center hemisphere, the points would be lined up vertically (or
horizontally) at 0 km/s.  Stars with no matching components are omitted.
\label{fig:2a} }
\end{figure}

The numerical quality of the fit of the AC and GC vectors to observed
component velocities in stars in the galactic hemisphere is tested by
plotting those components showing the smallest \dVi\ values ($\leq$1.3
\kms), for each star and each cloud.  Fig. \ref{fig:2a} shows a plot
of the minimum \dVi(AC) (x-axis) and the minimum \dVi(GC) (y-axis),
for a star i.  The \dVi\ values are determined after comparing
\dVi(AC) and \dVi(GC) calculated for each component seen towards the
star i \citep[for an earlier version of this plot see][]{Frisch:1997}.
If either the AC (or GC) vector perfectly described the velocity
components in the galactic center hemisphere, the points would be
lined up vertically (or horizontally) at 0 km/s. The scatter in \dVi\
values shows that neither velocity vector matches the data perfectly.
Many of the stars showing a GC component are distant (Fig. \ref{fig:1a}), where a
serendipitous coincidence in velocities is more likely.
The \dV\ standard deviations for the AC and GC components are,
respectively, 
0.7 \kms\ and 0.5 \kms, where only components used to derive the velocity vectors are
included.

\begin{figure}
\vspace*{4.in}
\includegraphics{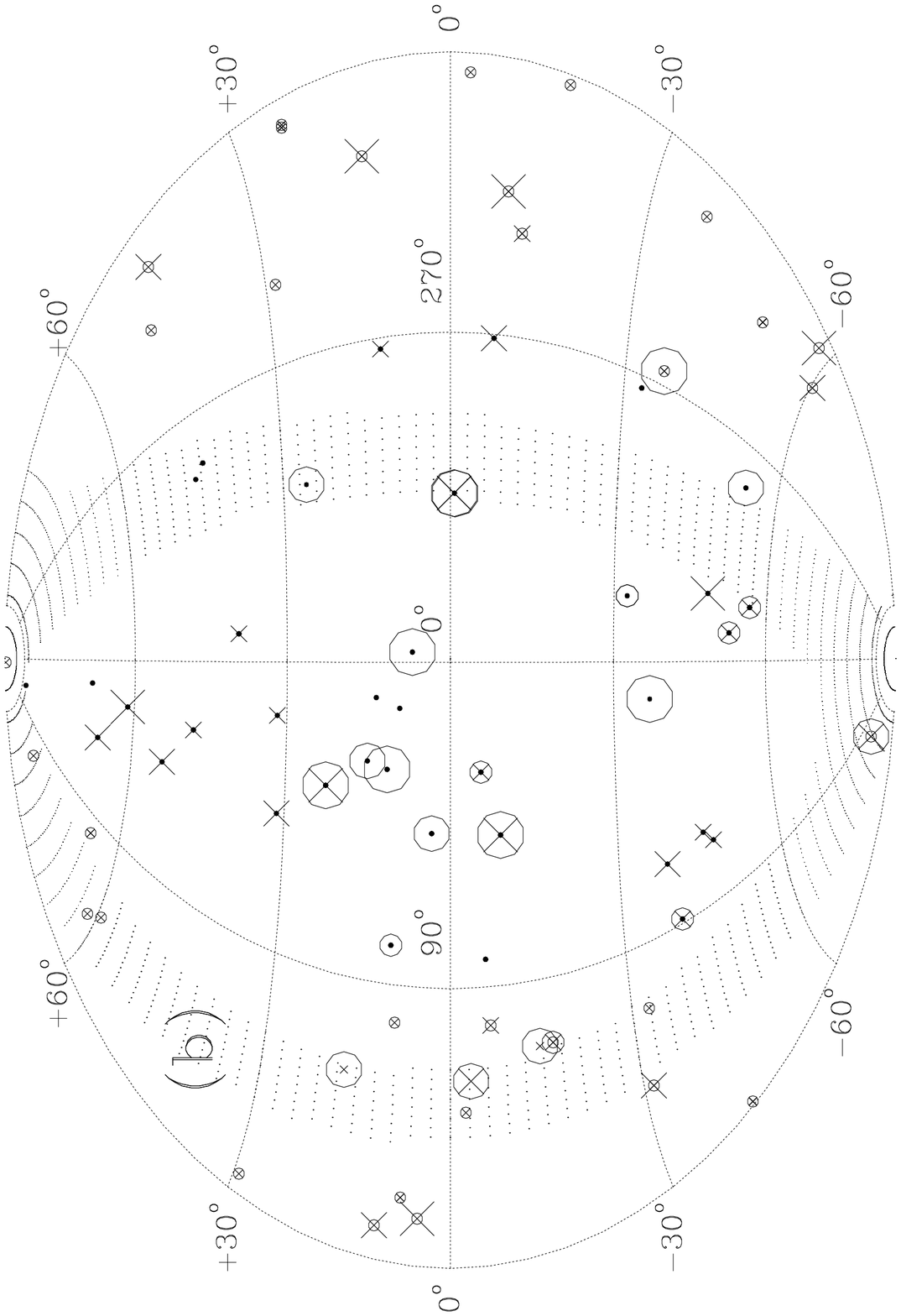}
\caption{ (b) Same as previous figure, except that the crosses are
components at the LIC velocity and the shading is the overlap between
the AC and LIC.  The components at the LIC velocity near
\glong$\sim$40\deeg, \glat=20--70\deeg\ appear to sample a tangential
region of the Upper Centaurus-Lupus shell of the Scorpius-Centaurus superbubble, 
which has expanded to the
solar vicinity \citep{Frisch:1981,Crawford:1991,deGeus:1992,Frisch:1995}.  The
nearest star in the group, HD 131156 (\glong=23\deeg, \glat=61\deeg),
is 7 pc away.
\label{fig:1b} }
\end{figure}

\begin{figure}
\vspace*{4.in}
\includegraphics{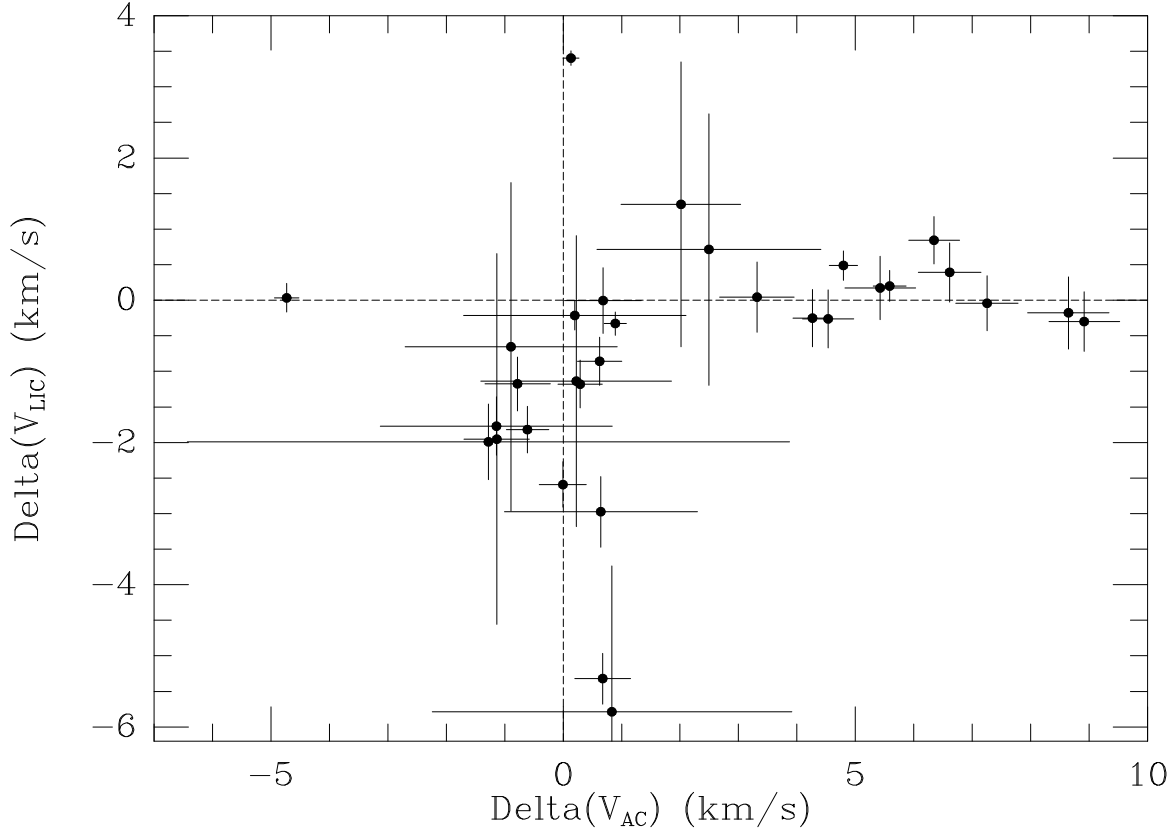}
\caption{(b) Same as previous figure, except that the vertical scale
displays components agreeing with the LIC velocity, for stars located
in the galactic center hemisphere.  The cluster of components at
\dVi(AC)$\sim$6 \kms, \dVi(LIC)$\sim$0 \kms\ appears to sample the
tangential region of the Upper Centaurus-Lupus shell, noted in
Fig. \ref{fig:1b}.  If this component group is actually the LIC, the
LSR upstream direction is towards \glong$\sim$346\deeg,
\glat$\sim$0\deeg.   Stars with no matching components are omitted.
\label{fig:2b} }
\end{figure}

For comparison, the LIC components in the galactic center hemisphere
are plotted in Fig. \ref{fig:2b}, where the y-axis is \dVi(LIC).  The
cluster of components at \dVi(AC)$\sim$6 \kms\ and \dVi(LIC)$\sim$0 \kms\
represent stars which sample the tangential region of the Upper Centaurus-Lupus 
superbubble (Fig. \ref{fig:1b}), and may coincidently have the LIC velocity.

If the AC is the closest upstream cloud to the Sun, then the relative
AC-Sun velocity $\sim$35 \kms, corresponding to $\sim$35 parsecs per
million years, suggests the AC is likely to replace the LIC as the
next interstellar cloud surrounding the solar system sometime within
the next $\sim 10^4$--10$^5$ years.  The velocities of the AC (--35.0
\kms) and bulk flow velocity vectors (--28.1 \kms) differ by
$\sim$25\%, but the directions of the AC and bulk flow velocity
vectors are within $\sim$8\deeg\ of each other.
In this case, in the rest frame of the LIC the AC velocity AC is
$\sim$10 \kms\ and is close to the sound velocity (8.3 \kms)
of a perfect gas at the LIC temperature ($\sim$6,300 K).
This coincidence suggests that the kinematical structure of the
nearest ISM may be related to sonic turbulence, and that a shock will
form where the AC and LIC clouds collide.  If the physical properties of the AC 
are similar to those of the LIC, the solar wind termination
shock distance in the nose direction should decrease by $\sim$25\%
(from simple equilibration of the solar wind and ISM ram pressures),
which is comparable to variations expected from the solar cycle
\citep[e.g.][]{Zank:1999,TanakaWashimi:1999,SchererFahr:2003}.  In
this case, the consequences for the inner heliosphere should be
negligible.  It is important to note in this context that $\sim$98\%
of the diffuse material in the heliosphere today is ISM, although the
inner heliosphere interplanetary medium is solar wind dominated
because of the $R^{-2}$ solar wind density dependence.  If the colder
$\alpha$ Pav component at the AC velocity is formed in the AC, the
boundary conditions of the heliosphere may differ substantially from current values.

If instead the GC is the closest upstream cloud to the Sun, then the strongest observed
nearby interstellar \CaII\ component, at the GC velocity towards $\alpha$ Oph,
indicates that dramatic changes are in store for the heliosphere when it enters the GC
within the next 10$^3$--10$^5$ years.
The ratio \NCaII/\NHI\ in the GC would be $<$7.1 x 10$^{-9}$, based on
observations of the GC towards $\alpha$ Cen 
\citep[][]{Crawford:1994,LinskyWood:1996}.  This gives \NHI$>$2.3 x 10$^{19}$ \cmtwo\
for the GC component towards $\alpha$ Oph.  This limit is close to
\NHI\ estimated for this component towards $\alpha$ Oph based on 21 cm data 
\citep[\NHI$\leq$3.6 x 10$^{19}$ \cmtwo][]{Frisch:1981}.  Since the GC \NHI\ towards
$\alpha$ Cen and 36 Oph are quite close, most of the GC density would be 
within 1.3 pc of the Sun,
implying a GC density of $\geq$5 \cc\ and 
future heliosphere shrinkage for possible cloud properties \citep{ZankFrisch:1999}. 

However the interpretation favored by this author, based on the above arguments 
and bias, is that the cloud order moving outwards from the Sun in the 
upstream direction consists of the LIC extending to virtually fill the $\alpha$ Cen
sightline (with the required velocity gradient or discontinuity, which is consistent
with the observed turbulence), followed by the AC which may or
may not extend to the 36 Oph sightline, and beyond that the higher column density
GC.  The absence of the LIC towards 36 Oph would indicate a filamentary structure
for the LIC.  In this case the $\alpha$ Pav cold component at the AC velocity would be
a serendipitous coincidence in velocities and unrelated to the AC.  
These derived cloud velocities, however, do not reflect these biases.
High-resolution high-signal-to-noise spectroscopy of additional sightlines, 
particularly in the UV,
are required to distinguish the correct interpretation of the data.

\acknowledgements 

This research has been supported by NASA grants NAG5-8163, NAG5-11999,
and NAG5-111005.


\clearpage
\input{table1}
\input{table2}

\clearpage

\clearpage

\clearpage

\clearpage

\end{document}

%% file: table1.tex
\begin{deluxetable}{llll llll c }
\tablecolumns{9} 
\tablewidth{0pc} 
\tablecaption{ISM Flows near the Sun\label{tab:vec}}
\tablehead{ 
\colhead{}&\multicolumn{3}{c}{HC Vector\tablenotemark{a}}&\colhead{}&\multicolumn{3}{c}{LSR Vector\tablenotemark{{\rm a,b}}}& \colhead{Notes} \\
\cline{2-4} \cline{6-8} \\ 
\colhead{}&\colhead{Vel.}&\colhead{l}&\colhead{b}&\colhead{}& \colhead{Vel} & \colhead{l}&\colhead{b} &  \colhead{} \\
\colhead{}&\colhead{(\kms)}&\colhead{(\deeg)}&\colhead{(\deeg)}&\colhead{}&\colhead{(\kms)}&\colhead{(\deeg)}&\colhead{(\deeg)} & \colhead{} \\
}
\startdata 
\hline 
\sidehead{Cloud surrounding solar system (LIC):}
$V$(He)&\HeV & \Hel & \Heb && --15.8 & 346.1 & 0.2 & \citet{Witteetal:2003} \\
& &  & &&  &  & & \citet{Flynne:1998} \\
\sidehead{Bulk flow of nearby ISM:}
$V$(96)&--28.1$\pm$4.6&12.4&11.6&$~~~$&--17.0&2.3&--5.2&60 stars \citep{Frischetal:2002} \\
\sidehead{``G'' Cloud:}
$V$(GC)& --29.1$\pm$0.5 & 5.3 & +19.6 && --17.7 & 351.2 & +8.5 & 18 stars (this paper) \\
$V$(GC)& --29.4$\pm$0.4         & 4.5 & +20.5 && --18.0 & 350.0 & +10.0 & \citet{LallementBertin:1992}\\
\sidehead{Apex Cloud:}
$V$(AC)&\ACV&\ACl&\ACb&$~~~$& --23.3$\pm$0.5 & 5.5 & 4.1 & 15 stars (this paper)\\
\hline 
\enddata 
\tablenotetext{a}{Galactic coordinates correspond to upstream directions.}
\tablenotetext{b}{Conversion to LSR uses the solar apex motion
($V$=13.38 \kms, towards galactic coordinates $l$=27.7$^o$, $b$=32.4$^o$) derived from 
Hipparcos data \citep{DehnenBinney:1998}.}

\end{deluxetable} 

%% file: table2.tex
\begin{deluxetable}{llcccc cc cc}
\tablecolumns{10} 
\tabletypesize{\small}
\tablewidth{0pc} 
\tablecaption{Stars in Galactic Center Hemisphere with Apex or G-Cloud Velocity Components\tablenotemark{a} \label{tab:AC}}
\tablehead{ 
\colhead{HD}& \colhead{Name}& \colhead{l}& \colhead{b}& \colhead{d}& \colhead{Spec.} &\multicolumn{2}{c}{AC Component} & \multicolumn{2}{c}{GC Component} \\
\cline{7-8} \cline{9-10} \\
\colhead{} & \colhead{} & \colhead{} & \colhead{} & \colhead{} & \colhead{} & 
\colhead{Velocity} & \colhead{\dV(AC)} & \colhead{Velocity}  & \colhead{\dV(GC)} \\
\colhead{}& \colhead{}& \colhead{\deeg}& \colhead{\deeg}& \colhead{pc} & \colhead{}& \colhead{\kms}& \colhead{\kms} & \colhead{\kms}  & \colhead{\kms} \\
}
\startdata 
128620 & $\alpha$ CenA &  315.7 &   --0.7 &   1.4 &  G2 V &  --17.8 & 0.6 & s & --0.1 \\
128621 & $\alpha$ CenB & 315.7 & --0.7 & 1.4 & K1 V&  --18.1 & 0.3  & s & --0.4 \\
209100  & $\epsilon$  Ind  &  336.2  &  --48.0  &    3.6  & K4.5V  &  \nodata & \nodata & --9.2  & --0.4 \\ 

155886/5  &   36 OphAB &   358.3   &   6.9  &   4.3 &  K0V/K1V  &  --35. : & --1.3 & --27.9  &  0.3 \\

187642 & $\alpha$ Aql &   47.7 &   --8.9 &   5.1 &   A7 V & --26.9 & --0.7 & --18.1 & 0.4 \\

165341 & 70 Oph & 29.9 & 11.4 & 5.1 &  K0V  &  -32.9  & 0.7 &  --26.6  & --0.2  \\


197481 & AU Mic & 12.7 & --36.8 & 9.9 & M0 & --21.2 & 0.7 & \nodata & \nodata \\

159561 & $\alpha$ Oph &   35.9 &   22.6 &   14.3 &   A5 III &  --32.0 & 0.2 &  --26.2   & --0.6 \\

36705 & AB Dor &  275.3 &  -33.1 &   14.9 &   K1III  & \nodata & \nodata  &  5.2 & --0.2 \\

115892 & $\iota$ Cen &  309.4 &   25.8 &   18.0 &   A2 V & --18.2 & --0.6 &  s & --0.1 \\
12311 & $\alpha$ Hyi &    289.5 &    --53.8 &     21.9 &     F0 V &  4.9 & 0.1 & s & 1.0 \\

177724 & $\zeta$ Aql &     46.9 &      3.3 &     25.5 &     A0 Vn & --29.7 & --1.1 & --20.6 & 0.5 \\
161868 & $\gamma$ Oph &     28.0 &     15.0 &     29.1 &     A0 V &   --33.1 &  0.9 & \nodata & \nodata \\

120418  & $\theta$  Peg  &   67.4  &  --38.7  &   29.6  &   A2IV  &   \nodata & \nodata & --4.2  & --0.3 \\

209952 & $\alpha$ Gru &    350.0 &    --52.4 &     31.1 &     B7I V &   --13.0 & --0.9 &  \nodata & \nodata \\
88955  &   HR   4023  &  274.3  &   11.9  &   31.5  &  A2V  &  \nodata & \nodata & --1.7 &  --0.2 \\

177756 & $\lambda$ Aql &     30.3 &     --5.5 &     38.4 &     B9 Vn &  --30.7 & 0.7 & \nodata & \nodata \\
218045 & $\alpha$ Peg &     88.3 &    --40.4 &     42.8 &     B9III & --0.5 &  0.2 & \nodata    & \nodata \\
135742  &  $\beta$  Lib  &  352.0  &   39.2  &   49.1  &  B8V  & \nodata & \nodata &  --26.9  &  0.0 \\

160613  &  $o$  Ser  &   13.3  &  9.2  &   51.5  &   A2Va  &  \nodata & \nodata &  --29.0 & --0.6 \\   
186882  & $\delta$  Cyg  &   78.7  &   10.2  &   52.4  &   B9.5IV  & \nodata  &  \nodata  & --9.6 & --0.1 \\ 
          
193924 & $\alpha$ Pav &    340.9 &    --32.5 &     56.2 &     B2 IV &  --19.6 &   0.0 &  \nodata & \nodata  \\ 
213998  &   $\eta$  Aqr  &   66.8  &  --47.6  &   56.3  &  B9IV--Vn  &  \nodata & \nodata & --2.1      &  --0.5 \\

156928  &  $\nu$  Ser  &   10.6  &   13.5  &   59.3  &   A0/A1V  & \nodata & \nodata & --27.7  &  1.2 \\

\enddata

\tablenotetext{a}{The velocities in columns 7 and 9 are the observed heliocentric component velocities.
\dV(AC) (\dV(GC) is the difference between the observed and projected velocities of $V$(AC)
($V$(GC)) in Table \ref{tab:vec}.
The velocity references are given in Paper I (\CaII) and RL02 (\MgII). 
An ``s'' indicates the components listed in columns 7 and 9 are the same. 
}
\end{deluxetable}